\documentclass[twocolumn,pre,a4paper]{revtex4}    
\usepackage{graphicx}
\newcommand{\ie}{{\em i.e. }}   
    
\newcommand{\cf}{{\em cf. }}

\begin{document}    
 LPT Orsay 03/73     
    
\title{\bf NETWORK TRANSITIVITY AND MATRIX MODELS}    
\author{Z. Burda $^1$, J. Jurkiewicz $^1$ and A. Krzywicki $^2$}    
\affiliation{$^1$ M.~Smoluchowski Institute of 
Physics, Jagellonian University,    
 ul. Reymonta 4, 30-059 Krakow, Poland\\    
$^2$ Laboratoire de Physique Th\'eorique, B\^atiment 210,     
Universit\'e Paris-Sud, 91405 Orsay, France\\}    
    
\begin{abstract}    
This paper is a step towards a 
systematic theory of the transitivity    
(clustering) phenomenon in random networks.
 A static framework is used,
with adjacency matrix playing the role 
of the dynamical variable. Hence,
our model is a matrix model, where 
matrices are random, but their 
elements take values 0 and 1 only. 
Confusion present in some papers 
where earlier attempts to incorporate 
transitivity in a similar 
framework have been made is hopefully 
dissipated. Inspired by more 
conventional matrix models, new 
analytic techniques to develop
a static model with non-trivial 
clustering are introduced. Computer 
simulations complete the analytic discussion.
    
\end{abstract}    
\maketitle    
    
\section{INTRODUCTION}    
Network model builders are currently adopting 
one of the two complementary    
approaches: Either a network is constructed 
step by step, by adding     
successive nodes and links. Or else, 
what is constructed is a static     
statistical ensemble of networks. Each 
of these two approaches has its     
merits and shortcomings: Evolving network 
models shed light on the growth     
dynamics, while static ensembles are more 
appropriate for the study of     
structural traits. The classical model 
of Erd\"os and R\'enyi \cite{bol} 
has been generalized so as to incorporate 
arbitrary degree distributions 
and even some correlations, but a serious 
shortcoming of the static models 
proposed so far is that they do not 
capture the common feature of most real 
networks: neighbors of a randomly chosen 
node are directly linked to each 
other much more frequently than by chance, 
so that many short loops appear. 
The networks tend to have locally 
a tree structure (see the reviews 
\cite{ab,dm,new}). And, as pointed 
out in ref. \cite{new}, {\em ``for 
general networks we currently have 
no idea how to incorporate transitivity 
into random graph models}''. In this 
paper we fill this 
gap, at least partially. The attention 
of the reader should be called to 
the very recent refs. \cite{doro,bp}, 
where the clustering problem is 
also addressed, but following very different avenues.   
\par    
Graphs are a mathematical representation 
of networks. For definiteness    
we consider in this paper undirected 
graphs only. Let us denote by $N$     
the number of nodes in a graph and by 
$M =\{M_{ij}\}, i,j=1,2,...,N$ the     
symmetric incidence matrix, with $M_{ii}=0$ 
on the diagonal and $M_{ij}=1$     
or $0$ depending on whether the nodes 
labeled $i$ and $j$ are connected     
or not. Whole information about a graph 
is encoded in its adjacency matrix.     
A general random graph model can be 
defined by introducing the partition     
function (see, for example, ref \cite{bl}):    
\begin{equation}\label{eq:pf}    
Z = \sum_M \; e^{S(M)}\, \delta\bigl( Tr(M^2) - 2L\bigr)    
\end{equation}    
where $L$ is the number of links and $S(M)$ is a function
which we will call the action. The sum is over    
all possible adjacency matrices $M$. The simplest choice 
is $S(M) = 0$. The corresponding graphs are those of 
the classical theory of Erd\"os and R\'enyi \cite{bol}: 
the value of the ratio $L/N$ determines a variant of the model.    
\par    
Probably the simplest extension of the 
classical theory consists 
in setting $S(M)=g Tr(M^3)$, 
directly proportional 
to the number of triangles. This has 
been attempted already 
many years ago by Strauss \cite{str}. 
His results are summarized 
in the recent review \cite{new}: {\em 
``There is     
however, one unfortunate pathology ... 
If, for example, we include a term     
in the Hamiltonian that is linear in 
the number of triangles in the graph,     
with an accompanying positive temperature 
favoring these triangles, then     
the model has the tendency to ``condense'', 
forming regions of the graph     
that are essentially complete cliques - 
subset of vertices within which     
every possible link exists... Network 
in the real world however do not     
seem to have this sort of ``clumpy'' transitivity''}.    
\par    
It appears to us that this negative 
conclusion, which faithfully reflects    
the content of ref. \cite{str}, is not 
quite right. There is nothing wrong in    
Strauss's work. However, it is very 
incomplete and due to this incompleteness    
unvoluntarily misleading. One of the 
aim of our paper is to give a fresh and    
comprehensive discussion of Strauss's model.    
\par     
The essence of Strauss's argument is 
as follows: assuming that the                    
ratio $L/N$ is kept constant, one can 
easily convince oneself that there     
exist pathological configurations for 
which $Tr(M^3) \propto N^{3/2}$.    
The contribution of such a configuration 
to the partition function is    
explosive in the large $N$ limit, since 
it cannot be tamed by the entropy    
factor falling roughly speaking like the 
inverse of the number of graphs,    
\ie like $\exp{(-\mbox{\rm const} \times N \log{N})}$. 
Thus, whatever small     
the coupling $g$ is, the only stable states 
of the system are the pathological 
ones, provided the system is large enough.     
\par    
As we will show later on, the pathological crumpled 
states - the Strauss phase - are    
separated from a smooth phase by a barrier 
that grows with increasing  
$N$. If the system is prepared in the smooth 
phase, it has a very tiny probability    
to roll out over the barrier to the Strauss 
phase. This probability tends rapidly    
to zero in the thermodynamic limit. Strauss 
has missed this point, because     
the systems he simulated were too small to 
signal the relative stability of    
the smooth phase. Now, for all practical 
purposes one can work in the smooth phase,    
ignoring the instability. This is what one 
does on many occasions in physics, in    
particular in the context of matrix models, 
where the instability also goes away    
when the matrix size tends to infinity.  
                                                                                   
\par    
We have mentioned matrix models on purpose. 
The theory of random matrices is an     
important branch of statistical physics, 
with applications ranging from nuclear    
physics to string theory. Some of the 
techniques developed in this theory can be    
adapted to a study of the model defined by 
(\ref{eq:pf}). This is also a matrix    
model, albeit dealing with rather special 
matrices: in standard matrix models    
the matrix elements are continuous random variables.    
\par    
The form (\ref{eq:pf}) of the partition 
function turns out to be very convenient    
for numerical simulations. In analytical 
calculations it will be convenient to    
use a slightly different formulation 
of the model, getting rid of the     
$\delta$-function and allowing small 
fluctuations of the number of links $L$.    
The partition function $Z$ will be, up to a factor,
the average of $\exp{(S)}$ in Erd\"os-R\'enyi theory. 
We first assume that a link is occupied with probability     
$p$. Hence, for given $N$ and $L$ the Erd\"os-R\'enyi weight is     
\begin{equation}\label{eq:w1}    
p^L (1-p)^{N(N-1)/2 - L} = (1-p)^{N(N-1)/2} 
(\frac{1}{p} - 1)^{-L}    
\end{equation}    
This primary weight is further multiplied 
by $\exp{(S)}$. Inserting (\ref{eq:w1}),    
integrating over $L$ and neglecting an 
irrelevant factor, we obtain the modified    
partition function    
\begin{equation}\label{eq:pf2}    
\tilde{Z} = \sum_M \exp{\bigl( - 
\frac{1}{2} \ln{(\frac{1}{p} -1)} Tr(M^2)    
+ S(M)\bigr)}    
\end{equation}    
In short, we have traded the $\delta$-function for a Gaussian. 
\par
In most of this paper we set 
$S(M)= g Tr(M^3)$, as in ref. \cite{str}. 
Thus, formally and up to a rescaling 
of the dynamical variable the model 
looks like the much studied matrix model    
\begin{equation}\label{eq:ex2}    
Z_{\mbox{\footnotesize \rm matrix}} = \int \text{d}M \; 
\exp{\bigl(-\frac{1}{2}Tr(M^2) + g Tr(M^3)\bigr)}    
\end{equation}    
where one integrates over all possible 
symmetric $N \times N$ matrices. The 
difference is in the integration measure, 
which is discrete in (\ref{eq:pf2})
and continuous            
\begin{equation}\label{eq:meas}    
\text{d}M = \prod_{i\leq j} dM_{ij}    
\end{equation}    
in (\ref{eq:ex2}). This difference is 
crucial, of course, but we would rather
like to insist on the similarities 
between the two models. In any case, the
example of the matrix model is for 
us a guide in our study. 
\par    
The plan of the paper is as follows: In 
Sect. II we develop $\exp{(S)}$ in
powers of $S$ and discuss the features of 
the perturbation series obtained
by integrating term by term. In Sect. 
IIA we recall how the behavior of the
perturbation series reflects the existence 
of an instability of the theory,
by considering two examples. In Sect. IIB 
we introduce a helpful diagrammatic
representation of the perturbative 
contributions to the partition function.
These diagrams are counted in Sect. IIC. 
It is argued that at finite $N$ the
perturbation series is pathological, 
indicating that nonperturbative
phenomena are in action. However, 
keeping only the terms that are non-vanishing
in the limit $N \to \infty$ one gets, 
like in the matrix model (\ref{eq:ex2}),
a convergent series. This series is summed 
in Sect. IID. We obtain a simple
analytic formula for the average number 
of triangles. We also show that the
introduction of the interaction $g Tr(M^3)$ 
leaves the degree distribution
unmodified. The nonperturbative dynamics 
is studied in detail in Sect. III,
using the Monte Carlo technique of 
numerical simulation. In a range of model
parameters we find a remarkable agreement 
between the data and the pertubative
predictions, showing that the nonperturbative 
phenomena are negligible in
this range. However, at large enough 
coupling strength the perturbation
theory breaks down, as expected. The 
transition point has an interesting 
scaling with $N$. This enables us to 
define the model so as to get a
non-trivial behavior of the clustering
 coefficient. In Sect. IV we
discuss possible generalizations. This section
contains also a summary of this 
work and a conclusion.
    
\section{PERTURBATION SERIES}    
\subsection{An analogy}    
Before entering into the main 
discussion of our problem let us consider    
an elementary example, to help 
those readers who are not conversant
with field theoretic arguments.    
\par    
Consider the following integral    
\begin{equation}\label{eq:ex1}    
I = \sqrt{\frac{\beta}{2\pi}} \int 
\text{d}x \; e^{\beta (-x^2/2 + g x^3/3    
- \epsilon x^4)}    
\end{equation}    
where $\epsilon$ is infinitesimal and 
has been introduced only in order to    
satisfy purists: $I$ can be regarded 
as the partition function of a particle    
subject to the combined action of a 
potential and of a heat bath. Formally,
the integrand resembles the summand 
in (\ref{eq:pf2}), except that the
integration variable is here just a number. 
\par
Consider a random walk in the potential 
given by the exponent in (\ref{eq:ex1}).
Assume that in some initial moment the 
particle is located at $x=0$. This 
is a metastable state. The particle 
eventually rolls over the barrier and    
reaches the deep minimum of the free 
energy at $x \approx g/4\epsilon$.    
As is well known, the lifetime $\tau$ 
of the metastable state is given by the    
Arrhenius formula~\cite{vk}:    
\begin{equation}\label{eq:arh}    
\tau \approx e^{\beta/6g^2}    
\end{equation}    
The decay of the metastable configuration 
is a nonperturbative phenomenon.    
The escape time has an essential 
singularity as a function of the coupling     
$g$. Of course, this nonperturbative 
phenomenon only occurs at nonzero    
temperature. When $\beta =\infty$ the 
particle stays forever in its initial    
position. Notice that the transition is 
more a cross-over than a genuine
phase transition. It occurs when the 
exponent in (\ref{eq:arh}) is of
order unity, but the value of $g$ where 
the transition occurs may slightly
depend on how the random walk is performed.  
\par    
In more complicated models the Arrhenius 
formula is not so readily derived.    
But nonperturbative dynamics shows up, if 
present, in the structure of the    
perturbation series in the coupling 
constant. Let us expand the exponential    
in (\ref{eq:ex1}) with respect to 
terms other than the quadratic one:    
\begin{equation}\label{eq:exp1}    
I = \frac{1}{\sqrt{\pi}}\sum_k      
\frac{\Gamma(3k+\frac{1}{2})}{\Gamma(2k+1)} \;    
\bigl(\frac{8 g^2}{9\beta}\bigr)^k \bigl( 1 + O(\epsilon)\bigr)    
\end{equation}    
It is evident that the series 
coefficients grow factorially and that the    
series has zero radius of convergence. 
This is a characteristic signal.    
We will find a similar behavior, at 
finite $N$, in the model defined by     
(\ref{eq:pf2}).    
\par
To conclude this section let us mention 
what happens when the partition
function instead of being a Riemann integral, 
like in (\ref{eq:ex1}), is
a matrix integral, like in (\ref{eq:ex2}): 
A perturbation expansion can 
again be defined and the terms in the 
expansion can be given a diagrammatic 
representation. A clever manner of 
cataloging these diagrams has been 
devised by 't Hooft \cite{thooft}. One first rescales 
$M \to \sqrt{N} M, g \to g/\sqrt{N}, 
\epsilon \to \epsilon/N$. One    
then observes that these diagrams can 
be drawn on a two-dimensional surface.    
Such a surface is always a sphere with 
a number of handles. Classes of diagrams    
are characterized by the number $h$ of these 
handles and the contributions of all    
diagrams belonging to the same class 
have the same $N$ dependence: $N^{2-2h}$.    
Summing over all $h$ one gets a badly 
divergent series. However, in the limit     
$N \to \infty$ the spherical topology 
(h=0) dominates and the corresponding     
series has a non-zero radius of convergence. 
We will seek a similar behavior
in our model. The hint is that one 
should carefully examine the 
$N \to \infty$ limit and that it 
may be wise to rescale the coupling 
constant in order to get a 
physically meaningful theory.    
    
\subsection{Diagrammatic representation}    
In this section we will introduce 
diagrams representing terms in the    
perturbative expansion of (\ref{eq:pf2}). 
To avoid misunderstanding let us    
stress from the outset that the diagrams 
introduced in this section are     
{\em not} to be identified with the 
graphs belonging to the statistical    
ensemble we are working with. These 
diagrams are just a tool helping to   
catalogue contributions to the 
partition function. We start by setting    
$S(M) = g Tr(M^3)$.
\par    
Let us expand in (\ref{eq:pf2}) the factor $e^S$:    
\begin{equation}\label{eq:exp2}    
\tilde{Z} = Z_0 \sum_n \frac{g^n}{n!} 
\left\langle [Tr(M^3)]^n\right\rangle_{ER}    
\end{equation}    
Here $Z_0$ is the partition function 
in the Erd\"os-R\'enyi ensemble of    
random graphs and the subscript ER in 
$\langle \dots\rangle _{ER}$ indicates that    
the average is calculated in this ensemble. 
Since $Z_0$ does not depend    
on our dynamical coupling $g$ it is 
for us an irrelevant normalization    
constant.    
\par    
The problem now is to calculate the 
averages appearing in the sum in    
(\ref{eq:exp2}). We use a method 
largely inspired by ref. \cite{bg},    
adopting also some of their notations, like    
\begin{equation}\label{eq:nota}    
N^{\underline{k}} = N!/(N-k)!    
\end{equation}    
which is the number of ways to 
choose $k$ among $N$ indices, the     
different permutations of the 
selected indices being considered as    
distinct. We have $Tr(M^3) = \sum_{abc} 
M_{ab} M_{bc} M_{ca}$, which    
is up to the factor $3!$ the number 
$T$ of triangles in the graph.    
We represent a matrix element $M_{ab}$ 
by a line segment. Indices  
$a,b$ are then associated with the 
ends of the segment. The product
$M_{ab} M_{bc} M_{ca}$ is represented 
by a triangle. Notice, that     
$M_{ab} M_{bc} M_{ca}$ is a random 
variable which can only take    
values $0$ or $1$. Since the diagonal 
elements of $M$ are by    
definition $0$ the probability that 
this product equals one is    
$p^3$. There are $N^{\underline{3}}$ 
configurations of indices    
which correspond to nonvanishing 
contributions to the sum. Hence    
\begin{equation}\label{eq:simplest}    
\langle Tr M^3\rangle_{ER} = 
p^3 N^{\underline{3}}    
\end{equation}    
The power of $p$ is equal to the 
number of the sides of the triangle    
and the underlined power of $N$ is 
the number of indices one    
is summing over. The contribution 
to the perturbation series    
is, of course, $g p^3 N^{\underline{3}}$.    
\par    
A complication arises as one goes to higher order of the    
perturbation theory. As one multiplies the traces one    
produces strings $M_{ab} M_{bc} \dots M_{ef}$ where some    
pairs of indices, possibly interchanged like $ab$ and $ba$,    
repeat themselves referring to the same element of the    
adjacency matrix. The corresponding probability factor is    
then $p$ and not some power of $p$. One has to identify such    
pairs of indices. It is also necessary to identify the     
independent summation indices and to count their    
number. One has also to count in how many distinct    
combinations the independent indices can appear in the    
string. All this may seem a bit confusing and is best    
explained with an example.    
\par
\begin{figure*}[htp] 
\includegraphics[width=14cm]{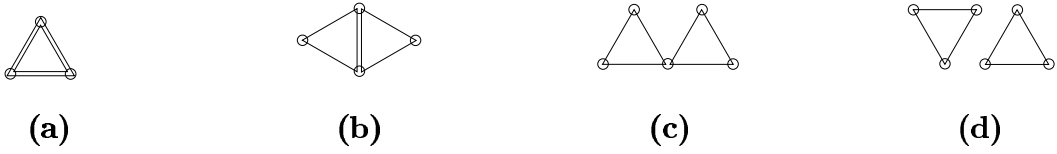}
\caption[Fig.1]{\label{fig1}
Diagrams representing $O(g^2)$ contributions 
to the partition function.} 
\end{figure*}

\begin{figure*}[htp] 
\includegraphics[width=14cm]{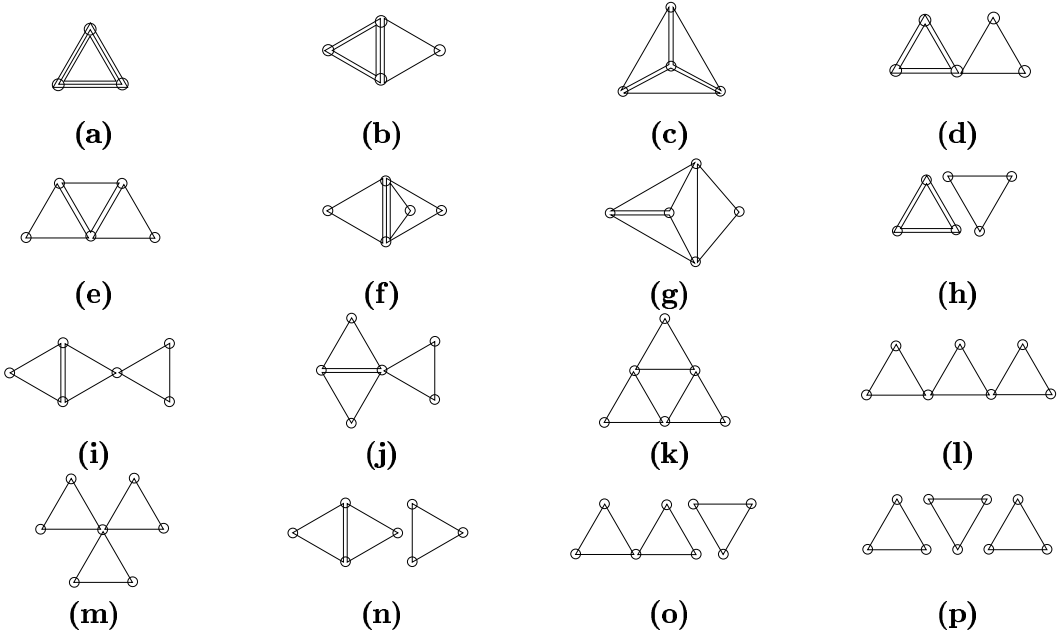}
\caption[Fig.2]{\label{fig2}
Diagrams representing $O(g^3)$ contributions 
to the partition function.} 
\end{figure*}
\par    
Consider the second order term in (\ref{eq:exp2}), $n=2$.    
One deals with the strings that have the following 
structure    
\begin{equation}\label{eq:string}    
M_{ab} M_{bc} M_{ca} M_{de} M_{ef} M_{fd}   
\end{equation}    
When one sums over indices there appear $N^{\underline{6}}$    
terms where the indices $abcdef$ are all distinct. Then,    
all the six index pairs are also distinct and the probability    
of individual strings is $p^6$. We illustrate this situation    
by drawing two triangles (because there are two traces) that    
are non-overlapping, like in Fig. 1d. The corresponding    
contribution to the perturbation series is     
$$ \mbox{\rm fig 1d} \rightarrow     
\frac{g^2}{2!} p^6 N^{\underline{6}}$$     
\par    
Another possibility is that the indices $def$ are identical to    
$abc$ but appear in a different order. For each choice of    
$abc$ there are $6$ such possibilities and there are     
$N^{\underline{3}}$ such choices. The probability of the    
string taking value $1$ is $p^3$. We illustrate this    
situation by drawing two overlapping triangles, as in Fig. 1a.    
Remember, that there are $6$ manners of putting one triangle    
on top of another. Hence    
$$ \mbox{\rm fig 1a} \rightarrow     
\frac{g^2}{2!} 6 p^3 N^{\underline{3}}$$     
Still another possibility is that two and only two indices are    
equal. They necessarily belong to two distinct traces. Thus    
one of the indices $def$ equals either $a$, or $b$, or $c$:    
there are three possible choices. Let us take one of them,    
say $a$. Then there are three possible structures for the    
second trace in the product: $M_{ae} M_{ef} M_{fa} , \; \;    
M_{ea} M_{af} M_{fe}$, and $M_{ef} M_{fa} M_{ae}$. Notice    
that $e$ and $f$ are dummy indices. We illustrate this situation     
by drawing two triangles with one common vertex. Remember that there    
are $3$ manners of attaching a triangle to a vertex of another    
triangle, like in Fig. 1c. On the whole there are $3 \times 3 = 9$     
arrangements of the five independent indices and six distinct index     
pairs. Thus    
$$ \mbox{\rm fig 1c} \rightarrow    
\frac{g^2}{2!} 9 p^6 N^{\underline{5}}$$    
Finally, two indices among $def$ can be identical to two indices    
among $abc$. There are three choices, let us take $ab$. Then there    
are six possible structures for the second trace in the product:    
$M_{ab} M_{be} M_{ea} , \; \; M_{ba} M_{ae} M_{eb} , \; \;     
M_{ae} M_{eb} M_{ba}$ ,  $M_{ea} M_{ab} M_{be} , \; \;     
M_{be} M_{ea} M_{ab}$ and  $M_{eb} M_{ba} M_{ae}$. 
We illustrate this situation    
by drawing two triangles with one common edge, 
like in Fig. 1b. Remember that    
attaching a triangle to two vertices can be 
done is six manners. There are    
$3 \times 6 = 18$ arrangements of the 4 
independent indices and five distinct    
pairs:    
$$ \mbox{\rm fig 1b} \rightarrow    
\frac{g^2}{2!} 18 p^5 N^{\underline{4}}$$    
The game can be extended to higher order $n$, 
although the number of diagrams    
increases very fast. The third order diagrams 
are listed in Fig. 2. The    
general rule is that the power of $p$ equals 
the number of triangle edges,      
\begin{center}
\begin{table}
\caption{Contributions to the partition 
function corresponding to
the diagrams of Fig. 2 ; the common factor 
$g^3/3!$ is omitted. \label{tab1}}
\vspace{0.4cm}
\begin{tabular}{|lr|lr|lr|lr|} \hline
    &  &  &  &  &  &  & \\
(a):& $36N^{\underline{3}}p^3\;$ &(b):& $324N^{\underline{4}}p^5\;$ &
(c):& $216N^{\underline{4}}p^6\;$ & (d):& $162N^{\underline{5}}p^6\;$ \\
    &  &  &  &  &  &  & \\ \hline
    &  &  &  &  &  &  & \\
(e):& $648N^{\underline{5}}p^7\;$ & (f):& $108N^{\underline{5}}p^7\;$ &
(g):& $324N^{\underline{5}}p^8\;$ & (h):& $18N^{\underline{6}}p^6\;$ \\
    &  &  &  &  &  &  & \\ \hline
    &  &  &  &  &  &  & \\
(i):& $324N^{\underline{6}}p^8\;$ & (j):& $324N^{\underline{6}}p^8\;$ &
(k):& $216N^{\underline{6}}p^9\;$ & (l):& $162N^{\underline{7}}p^9\;$ \\
    &  &  &  &  &  &  & \\ \hline
    &  &  &  &  &  &  & \\
(m):& $27N^{\underline{7}}p^9\;$ & (n):& $54N^{\underline{7}}p^8\;$ &
(o):& $27N^{\underline{8}}p^9\;$ & (p):& $1 N^{\underline{9}}p^9\;$  \\
    &  &  &  &  &  &  & \\ \hline
\end{tabular}
\end{table}
\end{center}
the number of triangle vertices appears 
as the underlined power of $N$ and,    
in the n$^{\mbox{\rm th}}$ order, there 
is a factor $g^n/n!$. The determination    
of the number of independent index 
arrangements is rather tedious. The    
best way is to proceed recursively.  
\par
Using the diagrams one can actually forget 
about indices. It is sufficient to    
construct the diagrams of 
order $n$ by adding one triangle to    
the diagrams of 
order $n-1$ in all possible manners. One has    
to multiply the number of arrangements factor in the target     
diagram of order $n-1$  by 
the number of ways the new triangle     
can be attached to it. These numerical 
factors should be added when a given    
diagram of order $n$ can be 
constructed from several     
diagram of order $n-1$. We 
repeat again the rules:    
\par    
- free triangle: factor 1    
\par    
- triangle attached to one vertex: factor 3    
\par    
- triangle attached to a pair of vertices: factor 6    
\par    
- triangle attached to three vertices: factor 6    
\par    
The general structure of the perturbation series is    
\begin{equation}\label{eq:exp3}    
\tilde{Z}/Z_0 = \sum_n \frac{g^n}{n!} \sum_k N^{\underline{k}}     
\sum_m W^{(n)}_{km} p^m    
\end{equation}   
The summation over $k$ goes from 
$3$ to $3n$. The power $m$ is always     
$\leq k$ and $m=k$ corresponds to 
diagrams where there is one or several    
groups of triangles lying one on 
top of another. Clearly, these are the    
only diagrams that survive in the 
limit $N \to \infty$.    
    
\subsection{Counting diagrams}    
The quantity $W^{(n)}_{km}$ appearing 
in (\ref{eq:exp3}) is the number of    
paths leading to a given diagram topology. 
In a sense it is the number of    
diagrams of that topology. We are able to 
determine it recursively, step by    
step, but we are unable to give a general 
formula for it. It is relatively    
easy to follow the evolution of the number 
of triangle vertices during the    
recursive process, it is much more tedious 
to keep track of the number of    
triangle edges \cite{foot0}. Thus, one can 
write a recursion equation for 
the sum    
\begin{equation}\label{eq:def}     
W^{(n)}_k = \sum_m W^{(n)}_{km}     
\end{equation}    
that is for the total number of diagrams 
of order $n$, with $k$ triangle    
vertices. This recursion relation, in 
essence, summarizes the rules    
listed in the preceding section:    
\begin{eqnarray}\label{eq:rec1}     
W^{(n+1)}_k = k(k-1)(k-2) W^{(n)}_k + \nonumber \\     
3(k-1)(k-2) W^{(n)}_{k-1} + 3(k-2) W^{(n)}_{k-2} + W^{(n)}_{k-3} 
\end{eqnarray}
The coefficients result from elementary combinatorics. The initial    
condition is $W^{(1)}_k = \delta_{k3}$. The first two iterations are
listed below. One can check that the numbers match those given in 
Figs. 1 and 2, provided the weights of diagrams with the same number 
of vertices are summed.
\par
{\bf $n=2$}
\begin{eqnarray}
W^{(2)}_3 & = &6 W^{(1)}_3 = 6 \nonumber \\
W^{(2)}_4 &=&18 W^{(1)}_3 = 18 \nonumber \\
W^{(2)}_5 &= &9 W^{(1)}_3= 9 \nonumber \\
W^{(2)}_6 &=& W^{(1)}_3 = 1 \nonumber 
\end{eqnarray}
{\bf $n=3$}
\begin{eqnarray}
W^{(3)}_3 &= &6 W^{(2)}_3 = 36 \nonumber \\
W^{(3)}_4 &= &24 W^{(2)}_4 + 18 W^{(2)}_3 = 540 \nonumber \\
W^{(3)}_5 &= &60 W^{(2)}_5 + 36 W^{(2)}_4 + 
9 W^{(2)}_3 = 1242 \nonumber \\
W^{(3)}_6 &= &120 W^{(2)}_6 + 60 W^{(2)}_5 + 
12 W^{(2)}_4 + W^{(2)}_3 = 882 \nonumber \\
W^{(3)}_7 &= &90 W^{(2)}_6 + 15 W^{(2)}_5 + 
W^{(2)}_4 = 243 \nonumber \\
W^{(3)}_8 &= &18 W^{(2)}_6 + W^{(2)}_5 = 27 \nonumber \\
W^{(3)}_9 &=& W^{(2)}_6 = 1 \nonumber 
\end{eqnarray}
One can easily see that the numbers given above agree with those
presented in Table \ref{tab1}. For example, the multiplicity
of the diagrams $(b)$ and $(c)$ is, respectively, $324$ and $216$, which
gives together $W^{(3)}_4=540$ as expected in the third order
for the sum of diagrams occupying $k=4$ vertices. 
\par
We can also estimate the number of
diagrams $W^{(n)}$ in the large order of the expansion,
that is for $n\rightarrow \infty$. As argued, this
number is expected to grow faster than a factorial, 
reflecting the non-perturbative transition to 
Strauss's phase. As we will 
show this is indeed the case.
Define the polynomial function    
\begin{equation}\label{eq:wn}     
W^{(n)}(y) = \sum_k  W^{(n)}_k y^k    
\end{equation}    
Multiplying both sides of eq. (\ref{eq:rec1}) by $y^{k-3}$ 
and summing over $k$ one obtains the following differential 
equation:    
\begin{equation}\label{eq:diff}     
W^{(n+1)}(y) = y^3 (\partial_y +1)^3 W^{(n)}(y)    
\end{equation}    
The solution is    
\begin{equation}\label{eq:sol}     
W^{(n)}(y) = [y^3 (\partial_y +1)^3]^n \cdot 1    
\end{equation}    
which can also be rewritten as    
\begin{equation}\label{eq:sol2}     
W^{(n)}(y) = e^{-y}[y^3 \partial_y^3]^n \cdot e^y    
\end{equation}    
Let us assume for a moment that 
$W^{(n)}(y)$ grows with $n$ less rapidly
than $(n!)^\kappa$, with some fixed $\kappa$, 
uniformly in $y$. It is then meaningful to introduce a
generating function
\begin{equation}\label{eq:sol3}
W(x,y) = \sum_n W^{(n)}(y)x^n/(n!)^\kappa
\end{equation} 
This function has a formal expansion
\begin{equation}\label{eq:sol4}
Z=e^{-y}{\cal W}\left(xy^3 \left(\partial^3_y\right)^3\right)e^y,    
\end{equation} 
where
\begin{equation}\label{eq:sol3a}
{\cal W}(z) = \sum_n \frac{z^n}{(n!)^\kappa}.
\end{equation}
Developing $e^y$ in a power series we have  
\begin{equation}\label{eq:sol5}
Z=e^{-y}\sum_k \frac{{\cal W}(x k (k-1)(k-2))y^k}{k!}.    
\end{equation}
To check the convergence of this 
sum we need the asymptotic
behavior of a function ${\cal W}(z)$. 
Examples of such functions for
integer $\kappa=0,1,2$ are well known, 
being a simple exponential $e^z$, 
the Bessel function $I_0(2\sqrt{z})$ 
or the generalized hypergeometric 
function (see \cite{wolf}) $_0F_2(z)$ 
respectively. For 
arbitrary $\kappa$ one has
\begin{equation}\label{asym}
{\cal W}(z) \sim 
\frac{C_\kappa}{z^\frac{\kappa-1}{2\kappa}}
e^{\kappa z^{1/\kappa}}\left(1+
\dots\right)
\end{equation}
with some $\kappa$-dependent constant $C_\kappa$. 
It is obvious that if $\kappa < 3$ the series (\ref{eq:sol5})
is meaningful only when $x=0$.
For $\kappa \ge 3$ it becomes convergent for arbitrary $y$ and $x$.
We conclude that $W^{(n)}$ grows faster than $(n!)^{3-\epsilon}$, 
but slower than $(n!)^3$ for arbitrary small $\epsilon$. Such an
explosive behavior of the number of perturbation theory diagrams
is a signal that nonperturbative phenomena are present. Indeed, 
it means that the coefficients of high powers of $g$ and $N^{-1}$ 
are increasing dramatically with the order of perturbation 
theory: what was assumed to be just a perturbation is in 
fact huge \cite{foot1} !

\subsection{Summation of leading diagrams}    
We are interested in the limit $N\rightarrow \infty$
with $pN = \alpha = {\rm const}$. The structure of the
perturbation expansion is given by eq. (\ref{eq:exp3}).
As already mentioned, in general, the number of triangle edges
(denoted $m$ in (\ref{eq:exp3})) is larger or equal to the
number of triangle vertices (denoted $k$ in (\ref{eq:exp3})).
In the limit under consideration, only
those diagrams contribute to the leading $N$-independent
term for which the number of triangle edges is equal to 
the number of triangle vertices. One can easily see that in these
diagrams the triangles can overlap, but otherwise do
not touch. In the expansion in $g$ up to the third order,
the following diagrams belong to this class:
a single triangle in the first order,
diagrams in Figs. 1a,d in the second,
and those in Figs. 3 a,h,p in the third.
Using the previously found results
we have up to the third order:
\begin{equation}
Z(G,\gamma) = 1 + \frac{G}{1!} \gamma + 
\frac{G^2}{2!} \left(\gamma + \gamma^2\right) +  
\frac{G^2}{3!} \left(\gamma + 3 \gamma^2 + \gamma^3 \right) +  
\dots 
\end{equation}
where the convenient notation $G=6g, 
\gamma=\alpha^3/6$ has been introduced.
In general, one can write this expansion as follows:
\begin{equation}\label{eq:expa}
Z(G,\gamma) =  1 + \sum_{n=1}^\infty 
\frac{G^n}{n!} Z^{(n)}(\gamma) =
1 + \sum_{n=1}^\infty \frac{G^n}{n!}
\sum_{k=1}^n z^{(n)}_k \gamma^k
\dots 
\end{equation}
The coefficients $z^{(n)}_k$ can 
be interpreted as the number of
all diagrams which consist of $n$ 
triangles located at $k$
isolated positions, with possible 
multi-occupation of a
position. Hence
\begin{equation}
z^{(n)}_k = \sum_P \frac{n!}{(n_1!)^{m_1}(n_2!)^{m_2}
 \dots (n_k!)^{m_k}
m_1! m_2! \dots m_k!}
\end{equation}
where $n_1 > n_2 > \dots n_k$ and the 
sum is over all the partitions 
$P$ of $n$: $n_1 m_1 + n_2 m_2 + \dots + n_k m_k = n$.
It turns out that the numbers $z^{(n)}_k$ 
can be calculated recursively:
\begin{equation}
z^{(n+1)}_k = k z^{(n)}_k + z^{(n)}_{k-1} \quad k=1,2,\dots,n+1
\end{equation}
with the initial condition: $z^{(1)}_1=1$ and  $z^{(n)}_k= 0$ for $k$
outside of the closed interval $[1,n]$. 
The meaning of the equation is as follows. If one adds a new triangle
to a configuration with $n$ triangles, this triangle
can be put at either of the $k$ existing positions or at a new position.
Hence, all configurations with $n+1$ triangles
located at $k$ positions can be obtained from 
configurations with $n$ triangles at $k$ positions, by placing a new
triangle at one of the $k$ old positions, or from configurations
with $n$ triangles at $(k-1)$ positions by placing a new
triangle at a new position.
The first few terms resulting from this recursion relation are:
\begin{eqnarray*}
Z^{(2)}(\gamma) & = & 1 \gamma^1 \\
Z^{(2)}(\gamma) & =  &1 \gamma^1 + 1 \gamma^2 \\
Z^{(3)}(\gamma) &= &1 \gamma^1 + 3 \gamma^2 + 1 \gamma^3 \\
Z^{(4)}(\gamma) &= &1 \gamma^1 + 7 \gamma^2 + 6 \gamma^3 + \gamma^4\\
Z^{(5)}(\gamma) &= &
1 \gamma^1 + 15 \gamma^2 + 25 \gamma^3 + 10 \gamma^4 + \gamma^5\\
Z^{(6)}(\gamma) &= &
1 \gamma^1 + 31 \gamma^2 + 90 \gamma^3 + 65 \gamma^4 + 
15 \gamma^5 + \gamma^6\\
\end{eqnarray*}
The recursion relation can be converted into a partial differential
equation for $Z(G,\gamma)$. Multiplying both sides of the equation
by $\gamma^{k-1}$ and summing over $k$ one finds
\begin{equation}
\frac{1}{\gamma} Z^{(n+1)} = \frac{\partial}{\partial \gamma} Z^{(n)}
 + Z^{(n)}
\end{equation}
where
\begin{equation}
Z^{(n)}(\gamma) = \sum_{k=1}^n z^{(n)}_k \gamma^k
\end{equation}
Now, multiplying both sides by $G^n/n!$ and summing over $n$ 
 one obtains
\begin{equation}
\frac{\partial}{\partial G} Z = 
\gamma \frac{\partial}{\partial\gamma}Z + \gamma Z
\end{equation}
where $Z$ is given by (\ref{eq:expa}). An even simpler equation is
satisfied by $F=\ln{Z}$:
\begin{equation}
\frac{\partial}{\partial G} F = 
\gamma \frac{\partial}{\partial\gamma}F + \gamma 
\end{equation}
One easily checks that the general solution is
\begin{equation}
F(\gamma,G) = f(\gamma e^G) -\gamma
\end{equation}
where $f$ is an arbitrary differentiable 
function. It results, however, from
(\ref{eq:expa}) that $F(\gamma,0)= 0$. Hence
\begin{equation}\label{eq:solF}
F(\gamma,G) = \ln{Z(\gamma,G)} = \gamma (e^G - 1)
\end{equation}
This result is not surprising for a 
practitioner of quantum field theory. Indeed,
$\ln{Z(\gamma,G)}$ should equal the sum of 
contributions of connected diagrams. The
only connected diagrams, in the large $N$ 
limit, are those where triangles are all
put on top of each other and according to 
our rules the diagram of $n$-th order
yields just $(g^n/n!) p^3 6^{n-1} 
N^{\underline{3}} \sim \gamma G^n/n!$.
\par    
Notice, that the same result (\ref{eq:solF}) 
is obtained assuming 
that in the Erd\"os-R\'enyi model the 
number of triangles has a Poisson 
distribution with average $\gamma$. Indeed, 
the average of $\exp{(GT)}$ is    
\begin{equation}\label{eq:poisson}     
\sum_{T=0}^{\infty} \frac{\gamma^T}{T!}e^{-\gamma} e^{GT} =     
\exp{[\gamma (e^G -1)]}    
\end{equation}    
The average number of triangles is
\begin{equation}\label{eq:Taver}     
\langle T \rangle =   \frac{\partial}{\partial G}\ln{Z}  
= \gamma e^G
\end{equation} 
It is important to note that in the 
$N \to \infty$ limit the average 
degree of a graph node becomes independent 
of $G$ and is just equal
to $\alpha$, like in pure Erd\"os-R\'enyi 
theory. This can be easily seen
in our formalism. Add to the action a source 
term $\eta Tr(PM)$, where
$P$ is the  matrix $P_{ij}=\delta_{i1}$, so 
that $Tr(PM)$ is the degree of 
the graph node with label 1. In our 
diagrammatics $\eta Tr(PM)$ produces
a line instead of a triangle. But only 
one end of this line has a running
index, the other end has index 1. The 
diagram of the lowest order in $\eta$
is just this line and gives the contribution 
$\eta p N = \eta \alpha$. 
All corrections due to the interaction 
$gTr(M^3)$ yield terms proportional
to some inverse powers of $N$, because 
there is no way to put a triangle
on a line. For example, the diagram of 
order $\eta g$, where one has one
triangle and one line on top of one of 
its edges gives 
$3\eta g p^3 N^{\underline{2}} \sim 3\eta g \alpha^3/N$ 
(we have 
$N^{\underline{2}}$ and not $N^{\underline{3}}$ 
because one of the triangle
edges has the fixed label 1). In conclusion: 
the only connected diagram of order
$O(\eta)$ is independent of $g$, as is, to 
this order the free energy $\ln{Z}$.
The average degree is just the derivative, 
at $\eta=0$, of the free energy
and equals $\alpha$. One can extend this 
argument to higher order moments
of the degree distribution.
\par
Using the results of this section we 
can propose a rough estimate of the 
expected region where nonperturbative 
physics sets in. With $p=\alpha/N$ 
and $N$ large the summand in (\ref{eq:pf2}) 
can be rewritten as     
\begin{equation}\label{eq:summand}     
\exp{\{\ln{\frac{N}{\alpha}} (-L + G_0 T)\}}    
\end{equation}    
We have rescaled the coupling by $\ln{\frac{N}{\alpha}}$, 
so that this large    
factor multiplies now both terms in the action. 
We expect that the perturbation    
series breaks down when the fluctuations of 
the two terms in the action become    
comparable. The number of links is $\sim N$ 
and we expect
$\langle(\delta L)^2\rangle \sim N$. In 
the large $N$ limit
the fluctuation of $T$ is
given by the second derivative of the 
free energy and equals
$\langle(\delta T]^2\rangle = \gamma \exp (G)
 = \gamma N^{G_0}$.
The two fluctuations become comparable when 
$G_0 \approx 1$. The
numerical calculations confirm the 
logarithmic scaling of $G$, but
as we will see in a moment the critical 
$G_0$ seems to lie 
below $1$.
\par

\begin{figure}
\includegraphics[width=6cm]{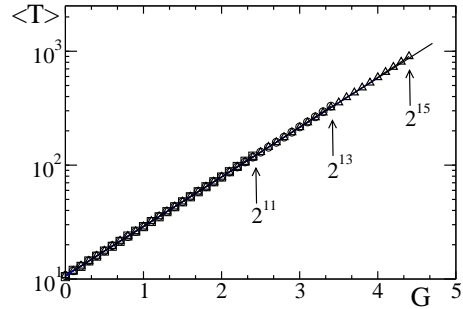}
\caption[Fig.3]{\label{fig3}
The average number of triangles $\langle T \rangle$ 
versus the coupling
constant $G \equiv 6g$. The average degree 
is set to $\alpha=4$ and
the simulation is performed for the number 
of nodes $N=2^{11}$ (squares), 
$2^{13}$ (circles) and $2^{15}$ (triangles).
 The arrows indicate the
position of the transition point $G=G_{out}$. 
The continuous line
represents the analytic result $\langle T \rangle = 
(\alpha^3/6)\exp{(G)}$.}.
\end{figure}

\section{NUMERICAL SIMULATIONS}    
We have shown in the preceding chapter 
that at finite $N$ the perturbative     
series has a behavior which signals the presence of a    
nonperturbative phenomenon. The 
similarity of our problem with the     
example exhibited in the first subsection 
suggests an educated guess:    
there is a barrier separating the 
perturbative phase from a pathological    
but stable configuration; the nonperturbative 
phenomenon in question is     
the rolling of the system over the barrier 
towards this stable configuration.     
The barrier must become unpenetrable in 
the $N \to \infty$ limit, because     
in this limit the perturbation series 
becomes well behaved, actually     
summable, whatever is the coupling. 
We will confirm this guess with the    
help of numerical simulations. 
\vspace{1cm}
    
\begin{figure}[htp] 
\includegraphics[width=7cm]{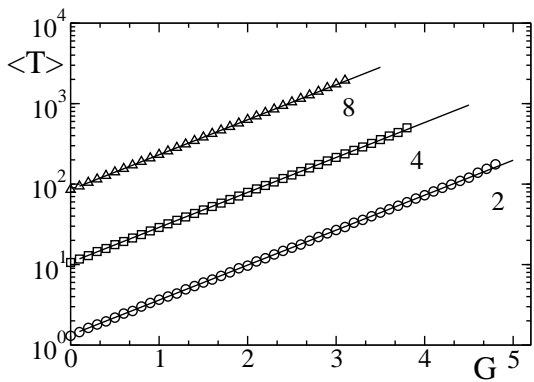}
\caption[Fig.4]{\label{fig4}
As in Fig. \ref{fig3}: $\langle T \rangle$ versus $G$, at 
$N=2^{14}$, but for three values of the average degree 
$\alpha=2$ (circles), $4$ (squares) and $8$ (triangles).} 
\end{figure}

In constructing an algorithm manipulating adjacency 
matrices it is most important to take into account 
the sparse nature of these matrices. Only the positions
of $2L$ matrix elements carry a relevant information. 
This makes it possible to reduce the amount of computer 
memory, needed to store an adjacency matrix,
from $O(N^2)$ in the naive coding to $O(N)$
in the linear coding. In effect, we simulate 
systems with the number of nodes of order $10^4$, \ie three orders
of magnitude larger then those simulated by Strauss \cite{str}. 
In the present work we use the algorithm introduced in refs. 
\cite{bck,bk} by straightforwardly upgrading it so as to include
the term in the action proportional to the number of triangles.
\par

\begin{figure}[htp] 
\includegraphics[width=7cm]{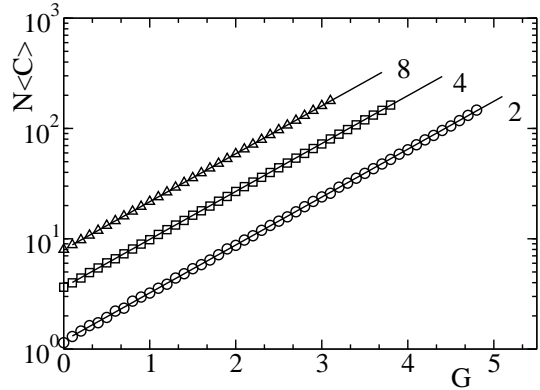}
\caption[Fig.5]{\label{fig5}
The scaled clustering coefficient $N \langle C \rangle$ versus $G$
at $N=2^{14}$ and for $\alpha=2$ (circles), $4$ (squares) and $8$ 
(triangles). The line represents the expected behavior 
$N \langle C \rangle \propto \exp{(G)}$, where the proportionality
coefficient is chosen so as to get the value expected in 
Erd\"os-R\'enyi model at $G=0$.} 
\end{figure}

In the first numerical experiment we set $\alpha = 2L/N = 4$ and we
measure the average number of triangles $\langle T \rangle$ for
$N = 2^{11}, 2^{13}$ and $2^{15}$. The coupling $G$ is changed in 
small steps until the system makes a transition to Strauss's phase.
The results are shown in Fig. \ref{fig3}. The continuous line 
corresponds to eq. (\ref{eq:Taver}). It is remarkable that the points
follow this line. The error bars are smaller than the symbol size.
The transition points are indicated by an arrow. A closer examination of the
data shows that near the transition the points start to deviate from the
line and lie systematically above it.
\vspace{1cm}

\begin{figure}[htp] 
\includegraphics[width=7cm]{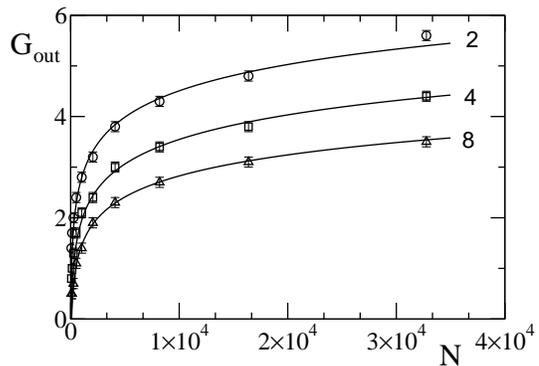}
\caption[Fig.6]{\label{fig6}
The transition from the perturbative to Strauss's phase occurs at
$G=G_{out}$. The figure shows how $G_{out}$ depends on the system size 
$N$, for $\alpha=2$ (circles), $4$ (squares) and $8$ (triangles). 
Notice the logarithmic growth of the curves.} 
\end{figure}

In the next experiment we set $N=2^{14}$ and measure $\langle T \rangle$
for $\alpha = 2, 4$ and $8$. The coupling $G$ again varies up to the
transition point. The result is shown in Fig. \ref{fig4}. The lines 
correspond to $\langle T \rangle = (\alpha/6)\exp{(G)}$.
The agreement is remarkable. We have also measured the local clustering 
measure $C_j$ as defined in ref. \cite{sw}:
\begin{equation}\label{eq:c}     
C_j = \frac{2 T_j}{L_j(L_j-1)}
\end{equation}  
where $T_j$ is the number of triangles touching the vertex $j$,
and $L_j$ is the number of links emerging from it.
We set $C_j=0$ when $L_j$ is zero or one. A global
clustering coefficient $C$ is obtained by averaging over vertices. In 
Fig. \ref{fig5} we plot $N \langle C \rangle$ versus $G$. It is seen 
that $N\langle C \rangle = \sigma (\alpha) \exp{(G)}$, with 
$\sigma (\alpha) = \alpha \bigl( 1 - (1 +\alpha ) \exp{(-\alpha )}\bigr)$,
which is the value of $N \langle C \rangle$ in Erd\"os-R\'enyi model. It is 
interesting that for very different values of $\alpha$ the transition 
occurs at roughly the same value of the clustering coefficient. 
This is presumably not a numerical accident, but we have no 
explanation to offer.
\vspace{1cm}

\begin{figure}[htp] 
\includegraphics[width=7cm]{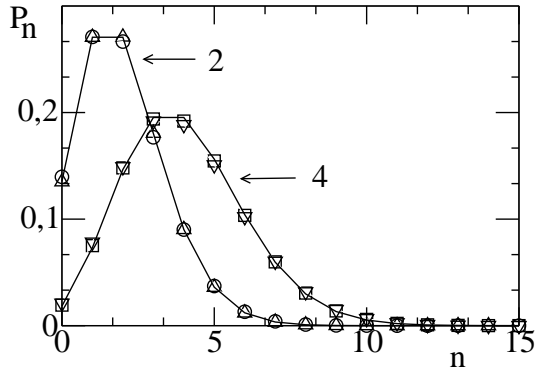}
\caption[Fig.7]{\label{fig7}
We compare the degree distribution calculated at two values of the
coupling constant $G$. The number of nodes is set to $N=2048$. 
At $\alpha=2$ the calculation is performed setting $G=0$ (triangles up) 
and $3.0$ (circles). At $\alpha=4$ the calculation corresponds to 
$G=0$ (squares) and $G=2.3$ (triangles down). The continuous lines
represent Poisson distributions with averages equal to $2$ and $4$,
respectively.} 
\end{figure} 
  
In Fig. \ref{fig6} we show the variation of the 
transition point $G=G_{out}$
with the system size $N$. In our experiments the 
coupling $G$ was always changed 
by $\delta G =0.1$. Thus, in the figure we have 
associated an error $0.1$ with
the data points. Here a comment is in order: 
After having changed $G$ we
always made $1000$ thermalization sweeps, then 
we carried out $20 000$ sweeps,
performing measures every 10 sweeps. It is 
important to remember that the number of
sweeps was always the same. Indeed, at {\em finite} 
$N$ the system will sooner or later
roll over the barrier, it is sufficient to wait 
long enough. The transition point 
$G=G_{out}$ is well defined when one decides 
to fix the waiting time. Actually,
we are more interested by the scaling of $G_{out}$ 
with $N$ than by its exact value.
\par
The curves in Fig. \ref{fig6} are
\begin{eqnarray}
G_{out} = 0.75 \ln{N} - 2.4 \;\; \mbox{\rm for}\;\; \alpha=2\\
G_{out} = 0.70 \ln{N} - 2.9 \;\; \mbox{\rm for} \;\;\alpha=4\\
G_{out} = 0.60 \ln{N} - 2.7 \;\; \mbox{\rm for}\;\; \alpha=8
\end{eqnarray}
It is very interesting, although not 
really surprising (see Sect. IID), that
$G_{out}$ scales like $\ln{N}$. This means 
that setting $G = G_0 \ln{N}$ one
obtains a model with the clustering 
coefficient scaling non-trivially:
$C \sim N^{G_0-1}$. 
\par
Fig. \ref{fig7} illustrates the fact 
that the degree distribution is in the
smooth phase insensitive to the value 
of the coupling $G$. We show the
distribution at $N=2048$ and $\alpha=2$ 
for $G=0$ and for a large value of $G$,
\ie $G=3.0$, close to $G_{out}$. The 
distributions are almost identical and
correspond to the Poisson distribution 
with average equal to $\alpha=2$ (the
line). This has been repeated for  $N=2048$ 
and $\alpha=4$, where we measured
at $G=0$ and $G=2.3$.

\section{DISCUSSION, SUMMARY AND CONCLUSION} 
\subsection{Possible generalizations}
Up to now we assumed that the interaction has a simple form
$S(M)=gTr(M^3)$. The question which immediately 
comes to mind is: what happens to
the network transitivity when the action is 
more complicated? Assume that $S(M)$
has the polynomial form: $S(M) = \sum_{n \geq 3} 
g_n Tr(M^n)$, with $g_3 \equiv g$.
Our diagrammatic rules can easily be extended 
to include this case. 
$Tr(M^n)$ is represented
by a polygon with $n$ sides. However the 
polygon can be folded, the same line segment  
being covered several times. In particular, 
when $n>4$, a polygon can be folded 
so that some of its edges form a triangle. 
\par
The calculation of leading diagrams 
given in Sect. IID can be 
generalized to interactions involving
 odd powers of $M$. One can limit
oneself to connected diagrams, those 
contributing to the free energy.
In the $N \to \infty$ limit 
the contribution of a diagram is 
proportional to $N$ to a power equal to 
\# vertices - \# edges (because $p=\alpha/N$). 
We are interested in diagrams
which overlap with a triangle. In this case 
\# vertices - \# edges $=0$. 
In all other cases this quantity is negative 
and the diagram does not contribute 
in the limit. It is easy to see that diagrams 
surviving in the $N \to \infty$ 
limit are those where triangles and folded 
polygons are put on top of each other.
\par
The situation is more complicated for even 
powers of $M$. The leading diagrams
look like branched polymers, with 
\# vertices = \# edges + 1, and their
contribution diverges like $N$. In 
order to avoid an unwanted renormalization
of the quadratic term in the action 
one has to subtract from $Tr(M^{2k})$ a
counterterm $\sim Tr(M^2)$ with an 
appropriate coefficient in front. Then the
calculation is like for the odd power case. 
\par
We have not pushed this calculation very far. 
As far as we can see one expects
a certain degree of universality: the higher 
powers of $M$ should not change
the qualitative picture very much, although 
they may be important for phenomenology,
to fit the data. A comprehensive study of 
these more general interactions is 
certainly worth being done. This is, however, 
beyond the scope of the present paper.
\par
It is not quite clear what is the best way 
of extending the theory of this paper
so as to obtain an arbitrary degree distribution. 
The field theoretical methods
extensively used in this paper usually fail 
when the action becomes nonanalytic.
The simplest, although perhaps not the most 
elegant, extension consists in using 
instead of the Erd\"os-R\'enyi model, a general 
model with uncorrelated nodes 
\cite{bk} as the zeroth order approximation. 
Preliminary numerical results
look encouraging, although it is clear that 
much has to be done in order to
get a fully satisfactory phenomenology. We 
hope to return to this problem
elsewhere. 

\subsection{Summary}
 Let us now summarize what has been 
achieved in this work: In most of this
paper we have discussed a model of 
random graphs where the classical 
Erd\"os-R\'enyi theory is generalized 
by the introduction into the action
of an interaction term $\frac{G}{3!}Tr(M^3)= GT$, 
$M$ being the $N\times N$ 
adjacency matrix and $T$ the number of triangles, 
respectively. This model 
is our guinea pig. It is a matrix model, but 
of a special kind, because 
the dynamical variable is a random matrix whose 
elements equal either 0 or 1. 
\par
Inspired by the analogy with more conventional 
matrix models we develop 
a diagrammatic technique, enabling us to 
calculate the perturbation series
analytically. We count the diagrams and 
show that, at finite $N$, their
number grows so rapidly that the perturbation 
series becomes pathological. 
This also happens in conventional matrix 
models and, as is 
well known, indicates the presence of a 
nonperturbative phenomenon. The 
nature of this phenomenon is identified 
through numerical simulations. 
There is a "potential barrier" and the  
system can roll over it and fall 
into a pathological phase, where all 
triangles form a unique clan. 
This phase was first discovered long 
ago by Strauss \cite{str}, who 
did not notice, however, that it is 
separated from a smooth 
phase by a barrier which becomes 
impenetrable at large $N$. We
propose to consider this smooth 
phase as the physical one.
\par
We show, that for large enough $N$ 
and in a range of values of the 
coupling constant $G$ the smooth 
phase can be considered,
for all practical purposes, as 
stable. In this range of $G$ it is
meaningful to neglect the nonperturbative 
physics and to limit oneself
to leading diagrams (those obtained setting
 $N=\infty$). We are able to
sum all these diagrams up, obtaining 
simple analytic expressions for
the free energy and for the average 
number of triangles. We also show
analytically that in this regime the 
degree distribution is insensitive 
to the value of the coupling $G$. 
\par
A heuristic argument, confirmed by 
numerical simulations, indicates that
the transition point $G=G_{out}$, where 
the system jumps to the Strauss's 
phase, scales with $N$ like $\ln{N}$. 
Hence, the physical coupling is not
so much $G$ but rather $G_0$ defined 
by the equation $G = G_0 \ln{N}$.
Our simulations indicate that at the 
transition point $G_0 = 0.6$ to $0.75$,
depending on the average degree, but 
this result should be taken with a
grain of salt. Anyhow, the clustering 
coefficient scales non-trivially,
like $C \sim N^{G_0-1}$ and is larger 
by one to two orders of magnitude
than in the unperturbed model.
\par
It appears that the analytic treatment 
can be extended to more complicated,
but polynomial actions. In the present 
state of affairs the extension of
our approach to more realistic, for 
example scale-free models can only
be done numerically. 

\subsection{Conclusion}
Clustering is a rather striking trait 
of many observed networks. The local
tree-like structure characterizing most 
static models is clearly non-realistic.
We have argued elsewhere that static models 
are an important ingredient of
network theory. Thus, we believe that it is 
important to be able to construct
static models with non-trivial clustering. 
There was some confusion concerning
the feasibility of such an enterprise. We 
hope having dissipated it. For the
sake of clarity we have focused our 
attention on a model where much can be done
analytically. It is a specific matrix model,
 where matrices are random,
but their elements take values 0 and 1 only. 
In the zero-th order approximation
it is equivalent to the classical Erd\"os-R\'enyi 
model of graphs. Non-trivial 
clustering is generated by an appropriate 
interaction. A comprehensive 
phenomenologically oriented study is 
beyond the scope of this paper
and remains to be carried out.

{\bf Acknowledgements:}

This work was partially supported by the EC IHP Grant     
No. HPRN-CT-1999-000161, by Polish State Committee for
Scientific Research (KBN) grant 2P03B 09622 (2002-2004), and by
EU IST Center of Excellence "COPIRA".  
Laboratoire de Physique Th\'eorique is Unit\'e Mixte     
du CNRS UMR 8627.


\begin{thebibliography}{99} 
\bibitem{bol} B. Bollob\'as, {\em Random Graphs},
(Academic Press, New York,     
2nd ed. 2001).   
\bibitem{ab} R. Albert, A.-L. Barabasi, Rev. Mod. 
Phys. {\bf 74}, 47 (2002).    
\bibitem{dm} S. N. Dorogovtsev, J.F.F. Mendes, 
Adv. Phys. {\bf 51}, 1079 (2002);    
{\em Evolution of Networks: from Biological 
Nets to to the Internet and WWW},     
(Oxford Univerity Press, New York, 2003).    
\bibitem{new} M.E.J. Newman, SIAM Review {\bf 45}, 167 (2003)  
\bibitem{doro} S. N. Dorogovtsev, cond-mat/0308444.
\bibitem{bp} M. Bogu\~na, R. Pastor-Satorras, cond-mat/0306072;
S. N. Dorogovtsev, cond-mat/0308336.
\bibitem{bl} J. Berg, M. L\"assig, Phys. Rev. Lett. 
{\bf 89},228701 (2002).       
\bibitem{str} D. Strauss, SIAM Review {\bf 28}, 
513 (1986) and references     
therein.    
\bibitem{vk} N.G. van Kampen, {\em Stochastic 
Processes in Physics and
Chemistry}, (North Holland, Amsterdam, 1992).    
\bibitem{thooft} G. 't Hooft, Nucl. Phys. {\bf72},461 (1974).    
\bibitem{bg} M. Bauer, O. Golinelli, J. Stat. Phys. 
{\bf 103}, 301 (2001).
\bibitem{foot0} We have also a recursive Monte 
Carlo algorithm
calculating $W^{(n)}_{km}$. The algorithm produces a 
string of indices, starting with 123 and adding 
successive triples. Suppose that at 
the $j$-th step, among $3j$ indices there are $i$
{\em distinct} ones. The algorithm goes over to 
the next step with probability
$P(i \to i+1) = {i+3 \choose 3}/{j+3 \choose 3}$. 
In the $(j+1)$-st step
three indices are drawn at random from the $i$ 
distinct indices supplemented
with three {\em indistinguishable} ones. These 
randomly selected indices are
accepted as the new triple with probability 
$1/6, 1/6, 1/3$ and $1$ when
they contain $3, 2, 1$ and $0$ indistinguishable 
indices, respectively.
After $n-1$ successfull steps one has 
constructed a string representing a particular contribution
to $Tr(M^n)$. One determines $k$ and $m$ by 
counting distinct indices 
and links. The acceptance of this algorithm 
drops rapidly with $n$, one 
can hardly go beyond $n=10$.   
\bibitem{wolf} http://functions.wolfram.com/HypergeometricFunctions/
\-HypergeometricPFQRegularized/06/02/04/ 
\bibitem{foot1} In field theory such a 
behavior of the number of diagrams
implies that the radius of convergence 
of the perturbation series is zero.
This is not the case in our model. The 
terms of the series are the largest
when $p=1$. But for this value of $p$ 
all links of the graph are, by 
construction, occupied and the sum of 
all diagrams of order $n$ is just 
$[N^{\underline{3}}]^n$ (\cf ref. \cite{bg}). 
Consequently the sum of
the series (\ref{eq:exp3})) is finite 
and less than 
$\exp{(gN^{\underline{3}})}$. The 
series is pathological, but it is
regulated as a result of a specific 
excluded volume effect. Indeed 
$N^{\underline{k}}=0$ for $k>N$, so 
that for fixed $N$ the
proliferation of diagrams does not 
continue indefinitely.
\bibitem{sw} S.H. Strogatz, D.J. 
Watts, Nature, {\bf 393}, 440 (1998).
\bibitem{bck}  Z. Burda, J.D. Correia, A. Krzywicki, 
Phys. Rev. E {\bf 64}, 046118 (2001).
\bibitem{bk} Z. Burda, A. Krzywicki, Phys.
 Rev. E {\bf 67 }, 046118 (2003);
The distinctive trait of this paper, 
compared to other publications on the  
subject, is that it deals with simple 
graphs, without self- and 
multiple-connections between nodes 
and contains a discussion of important 
finite $N$ effects, absent in pseudo-graphs.
\end{thebibliography}
\end{document}